\def\year{2016}\relax
\documentclass[letterpaper]{article}
\usepackage{aaai16}
\usepackage{times}
\usepackage{helvet}
\usepackage{courier}
\usepackage{float}
\usepackage{graphicx}
\usepackage{url}
\graphicspath{{figure/}}
\newenvironment{nfig}[3]{
	\begin{figure}[H]	
		\centering \includegraphics[width=#1\columnwidth]{#2}
		\caption{#3}\label{fig:#2}}
	{\end{figure}}
\newcommand{\fref}[1]{Figure~\ref{#1}}

\begin{document}
\title{Incentive Engineering Framework for Crowdsourcing Systems}
\author{Nhat V.Q. Truong, Sebastian Stein, Long Tran-Thanh \\
	School of Electronics and Computer Science \\ 
	University of Southampton \\ Southampton, UK \\ 
	\textit{\{nvqt1r14, ss2, ltt08r\}@ecs.soton.ac.uk} \\
	\And 
	Nicholas R. Jennings \\ 
	Department of Computing \\ 
	Imperial College London \\ London, UK \\ 
	\textit{n.jennings@imperial.ac.uk}}

\maketitle
\begin{abstract}
	Significant effort has been made to understand user motivation and to elicit user participation in crowdsourcing systems. However, incentive engineering, i.e., designing incentives that can purposefully motivate users, is still an open question and remains one of the key challenges of crowdsourcing initiatives. In this work in progress, we propose a general and systematic incentive engineering framework that system designers can use to implement appropriate incentives in order to effect desirable user behaviours.
\end{abstract}

\section{Introduction}

In crowdsourcing applications, motivating the crowd plays a crucial role in ensuring the success of the systems \cite{hossain12_crowdsourcing}. This is because the crowd is the key component that creates value in any crowdsourcing system. Specifically, most systems depend critically on large numbers of tasks being performed by the members of its crowd. However, motivating the crowd is also one of the most challenging issues facing crowdsourcing initiatives. Furthermore, it needs to be addressed throughout the life cycle of the initiative, from first attracting a crowd to the system, then encouraging their first contribution, to keeping them engaged in the longer term \cite{simula13_rise}.

To achieve this, system designers can provide a wide range of incentive mechanisms. These have the potential to enhance the users' existing motivations and thereby to align their behaviour with the system designers' goals. There is a significant body of research about user motivations, incentive mechanisms, as well as user behaviours in crowdsourcing (Kaufmann et al. \citeyear{kaufmann11_more}; Scekic et al. \citeyear{scekic13_incentives}, Mason and Watts \citeyear{mason10_financial}). However, existing work typically focuses on the motivations of users in specific systems, considers specific incentives, or specific behaviours of the crowd. In order to deliver appropriate incentives, many related aspects of the problem (i.e., behaviours, motivations, and incentives) should be considered. Hence, there is need for a more \textit{complete} approach that covers all necessary aspects in every crowdsourcing system. This will help give system designers an overall view about the incentive problem and its available solutions so that they can make appropriate choices (i.e., incentive mechanisms) over different time periods of their systems.
Some research has already started to address the problem more broadly by building models or frameworks that cover many cases or that can be used in many systems (Allahbakhsh et al. \citeyear{allahbakhsh13_quality}; Tokarchuk et al. \citeyear{tokarchuk12_analyzing}). However, these studies do not provide a clear process to be used in each stage of the life cycle of real systems. Hence, designers need to have a \textit{systematic} approach to the incentive problem that can help them come up with reasonable incentives by following concrete steps in the analysis and design process.

To meet these requirements, we have built a novel \textit{incentive engineering framework}, which provides a \textit{complete} and \textit{systematic} approach to delivering appropriate incentives to motivate contributors and align their behaviours in a way that is beneficial to the systems. In the next section, we briefly introduce our incentive engineering framework. Full details of the framework and its usage are presented at \url{http://i-crowdsourcing.com}.

\section{Incentive Engineering Framework}

Based on the above-mentioned idea about the connections between user behaviours, user motivations and incentive mechanisms, we have designed the following structure of the framework (\fref{fig:dimensions}) with three dimensions (corresponding to these three concepts): population aspects (population for short), design objectives, and actions. The connections between these dimensions are called action-design objective connections (or AO connections) and design objective-population connections (or OP connections).

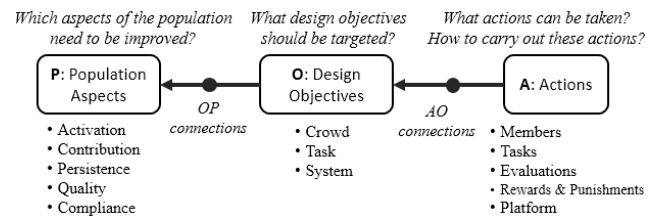
\begin{nfig}{1}{dimensions}{Three-Dimensional Structure of the Framework}\end{nfig}

Regarding the population aspects, we adapted three important processes associated with human motivation from Arnold et al. (\citeyear{arnold10_work}) (i.e., direction, persistence, and intensity) as three aspects: \textit{activation} (the extent to which new users join the system), \textit{persistence} (the extent to which users perform tasks regularly or intensively), and \textit{quality} (how well users perform tasks). Additionally, one dominant issue in crowdsourcing is malicious workers and spammers, that can decrease quality and hence affect the success of the whole system. We put this as the fourth aspects: \textit{compliance} (the extent to which users comply with system regulations and rules). And finally, from the workers' perspective, the first tasks are usually unfamiliar and more difficult than others. Therefore, completing these tasks is particularly challenging, but then future tasks are significantly easier. Therefore, we added \textit{contribution} (the extent to which users conduct the first tasks after becoming members of the system) as an important aspect.

Regarding the design objectives, based on motivation theories, we have identified over twenty system characteristics that have a significant influence on user motivation so that they can be used to align user behaviours effectively. These characteristics are grouped into three categories corresponding to three sources of motivation: \textit{crowd} (related to size, structure and interaction of the worker community), \textit{task} (related to designing tasks) and \textit{system} (related to the platform and system goals). For example, \textit{task autonomy} (a design objective in group \textit{task}) is the extent to which the system provides freedom to workers in choosing and performing tasks.

And finally, we have collected from the literature of economics, psychology, sociology, etc. over twenty actions that designers can consider to use to align user behaviours effectively. These actions are grouped into five categories: \textit{members} (related to individual workers or groups of them), \textit{tasks} (related to designing and delivering tasks), \textit{evaluations} (related to assessment methods of user performance), \textit{rewards \& punishments} (related to rewarding or punishing based on user activities) and \textit{platform} (related to features provided by the system). For example, \textit{using relative evaluation} (belonging to group \textit{evaluation}) is an action whereby the system evaluates the performance of users based on comparing their work with that of others in a specified group. Due to restricted space, these design objectives and actions cannot be described here but can be found on our website.

\begin{nfig}{1}{itable}{Simplified Version of the Incentive Table}\end{nfig}

Based on this three-dimensional structure, a table (called \textit{incentive table}) was built to aid the analysis process in a visual manner. It displays \textit{dimension items}\footnote{A dimension item can be a population aspect, a design objective or an action.} and their connections in a double-table structure. An abstract illustration of the general structure is shown in \fref{fig:itable}. The left table (OP table) has population aspects as columns and design objectives as rows. Similarly, columns of the right table (AO table) are specific actions that designers can use to regulate design objectives on the rows.

Between two dimension items, there can be a connection (AO or OP connection) or not. These connections can be positive or negative (shown as plus or minus signs). For example, design objective \textit{task autonomy} has a positive effect on population aspect \textit{persistence}. That means contributors will be motivated to perform tasks regularly when they have enough freedom to choose and perform tasks.
There is evidence in the literature to support the existence of some relationships. For example, \citeauthor{chen11_opportunities} (\citeyear{chen11_opportunities}) support the positive effect of design objective \textit{task clarity} on population aspect \textit{quality} by demonstrating that clear and well-designed task instructions improve performance results significantly. These are shown as shaded cells. However, some others are currently left as hypotheses and these are shown as unshaded cells.

\subsection{Usage of the Framework}

Users of the framework can be system designers (in building and maintaining their systems) or researchers (in understanding connections between dimension items, i.e., AO/OP connections). The table can be used in two different ways: \textit{backward} (to understand user behaviours with current incentive mechanisms) and \textit{forward} (to deliver new appropriate incentives after understanding reasons for user activities and wanting to improve some of these behaviours).

The \textit{backward} process starts from established actions, goes backward to see what system characteristics (i.e., design objectives) are affected by these actions, then continues to go backward to comprehend the total effects of these characteristics on the population. Hence, going backward can help designers identify problematic population aspects and the reasons. 
The \textit{forward} process starts from the population aspects that need to be enhanced, goes forward to choose what design objectives need to be focused on to influence on these aspects, and then continues going forward towards the AO table to decide what actions need to be changed so that they can achieve the desired outcomes on target design objectives. Therefore, going forward can help designers introduce new incentives to improve certain behaviours of users.

\section{Conclusion}

To deliver appropriate incentives to align user behaviours, we need to consider a broad range of available mechanisms and their effects on user motivation. Our framework provides a general approach for this incentive problem. As future work, we aim to increase the framework's applicability by investigating the connections between some of key dimension items, i.e., AO and OP connections, in more detail.

\makeatletter
\renewcommand\@biblabel[1]{}
\makeatother

\bibliographystyle{aaai}
\bibliography{truong}

\begin{thebibliography}{}

\bibitem[\protect\citeauthoryear{Allahbakhsh \bgroup et al\mbox.\egroup
  }{2013}]{allahbakhsh13_quality}
Allahbakhsh, M.; Benatallah, B.; Ignjatovic, A.; Motahari-Nezhad, H.~R.;
  Bertino, E.; and Dustdar, S.
\newblock 2013.
\newblock Quality control in crowdsourcing systems: {{Issues}} and directions.
\newblock {\em IEEE Internet Computing} 17(2):76--81.

\bibitem[\protect\citeauthoryear{Arnold, Randall, and
  Patterson}{2010}]{arnold10_work}
Arnold, J.; Randall, R.; and Patterson, F.
\newblock 2010.
\newblock {\em Work psychology: understanding human behaviour in the
  workplace}.
\newblock Harlow: {Financial Times Prentice Hall}, 5th edition.

\bibitem[\protect\citeauthoryear{Chen \bgroup et al\mbox.\egroup
  }{2011}]{chen11_opportunities}
Chen, J.~J.; Menezes, N.~J.; Bradley, A.~D.; and North, T.~A.
\newblock 2011.
\newblock Opportunities for crowdsourcing research on {{Amazon Mechanical
  Turk}}.
\newblock {\em Interfaces} 5(3).

\bibitem[\protect\citeauthoryear{Hossain}{2012}]{hossain12_crowdsourcing}
Hossain, M.
\newblock 2012.
\newblock Crowdsourcing: activities, incentives and users' motivations to
  participate.
\newblock In {\em Innovation {{Management}} and {{Technology Research}}
  ({{ICIMTR}}), 2012 {{International Conference}} on},  501--506.
\newblock {IEEE}.

\bibitem[\protect\citeauthoryear{Kaufmann, Schulze, and
  Veit}{2011}]{kaufmann11_more}
Kaufmann, N.; Schulze, T.; and Veit, D.
\newblock 2011.
\newblock More than fun and money. {{Worker}} motivation in crowdsourcing --
  {{A}} study on {{Mechanical Turk}}.
\newblock In {\em {{AMCIS}}}, volume~11,  1--11.

\bibitem[\protect\citeauthoryear{Mason and Watts}{2010}]{mason10_financial}
Mason, W., and Watts, D.~J.
\newblock 2010.
\newblock Financial incentives and the ``performance of crowds''.
\newblock {\em ACM SigKDD Explorations Newsletter} 11(2):100--108.

\bibitem[\protect\citeauthoryear{Scekic, Truong, and
  Dustdar}{2013}]{scekic13_incentives}
Scekic, O.; Truong, H.-L.; and Dustdar, S.
\newblock 2013.
\newblock Incentives and rewarding in social computing.
\newblock {\em Communications of the ACM} 56(6):72--82.

\bibitem[\protect\citeauthoryear{Simula}{2013}]{simula13_rise}
Simula, H.
\newblock 2013.
\newblock The rise and fall of crowdsourcing?
\newblock In {\em System {{Sciences}} ({{HICSS}}), 2013 46th {{Hawaii
  International Conference}} on},  2783--2791.
\newblock {IEEE}.

\bibitem[\protect\citeauthoryear{Tokarchuk, Cuel, and
  Zamarian}{2012}]{tokarchuk12_analyzing}
Tokarchuk, O.; Cuel, R.; and Zamarian, M.
\newblock 2012.
\newblock Analyzing crowd labor and designing incentives for humans in the
  loop.
\newblock {\em IEEE Internet Computing} 16(5):45--51.

\end{thebibliography}
\end{document}